\documentclass[%
 aip,
 jmp,
 amsmath,amssymb,
preprint,%
]{revtex4-1}

\usepackage{graphicx}
\usepackage{bm}
\usepackage{amsthm}
\usepackage{stmaryrd}
\usepackage{amsfonts}

\newcommand{\de}{\partial}
\newcommand{\ep}{\varepsilon}
\newcommand{\frin}[3]{{\left( \frac{#1}{#2} \right)}^{#3}}
\newcommand{\rf}[1]{(\ref{#1})} 
\renewcommand{\Gamma}{\varGamma}
\renewcommand{\Lambda}{\varLambda}
\renewcommand{\gamma}{\ep}
\renewcommand{\Psi}{\varPsi}
\renewcommand{\Phi}{\,\varPhi}
\newcommand{\nonu}{\nonumber}

\newcommand{\dst}{\displaystyle}

\newcommand{\Bs}{\bm \gamma}
\newcommand{\E}{\bm E}
\newcommand{\BE}{\bm E}
\newcommand{\Bj}{\bm D}
\newcommand{\bep}{\bm \varepsilon}

\newcommand{\I}{\bm I}
\newcommand{\J}{\left[\begin{array}{cc}
                 1 \\[-1ex]
                 i
                \end{array}
                \right]}

\newcommand{\M}{{\bm M}}
\newcommand{\re}{\mbox{Re\,}}
\newcommand{\im}{\mbox{Im\,}}
\newcommand{\om}{\omega}
\newcommand{\omo}{\tau_1}
\newcommand{\omt}{\tau_2}
\newcommand{\te}{\vartheta}

\newcommand{\ui}{u_{in}}
\newcommand{\ue}{u_{ex}}

\newcommand{\se}{\gamma_{ex}}
\newcommand{\ga}{\gamma}
\renewcommand{\ge}{\gamma_{ex}}
\newcommand{\gi}{\gamma_{in}}

\renewcommand{\r}{{\bm r}}

\renewcommand{\Pi}{\Phi_{in}}

\renewcommand{\P}{P_{m,n}}
\renewcommand{\leq}{\leqslant}
\renewcommand{\geq}{\geqslant}
\renewcommand{\iint}{\int}

\newcommand{\pr}{\prime}
\newcommand{\prr}{{\prime \prime}}
\newcommand{\la}{\langle}
\newcommand{\ra}{\rangle}

\renewcommand{\phi}{\varphi}
\renewcommand{\theequation}{\mbox{\arabic{section}.\arabic{equation}}}

\begin{document}

\title{
Effective complex permittivity tensor of a periodic array of cylinders
}
\author{Yuri A. Godin}
\affiliation{
$^1$Department of Mathematics and Statistics,
University of North Carolina at Charlotte,
Charlotte, NC 28223, U.S.A.}
\email{ygodin@uncc.edu}

\date{\today}

\begin{abstract}
We determine the effective complex permittivity of a two-dimensional composite,
consisting of an arbitrary doubly periodic array of identical circular cylinders in a homogeneous matrix, and whose dielectric properties are complex-valued. Efficient formulas are provided to determine the effective complex permittivity tensor which are in excellent agreement with numerical calculations. We also show that in contrast to the real-valued case, the real and imaginary parts of the effective complex-valued tensor can exhibit non-monotonic behavior as functions of volume fraction of cylinders, and can be either greater or less than that of the constituents. 

\end{abstract}


\maketitle

\section{Introduction}
\setcounter{equation}{0}
\renewcommand{\theequation}{\arabic{equation}}

The goal of this work is to find the effective permittivity tensor for an {\it arbitrary} doubly periodic array of circular cylinders when dielectric properties of the cylinders and the host medium are complex-valued. 
The problem of determining effective properties of a periodic array of inclusions was first considered by Rayleigh \cite{R:92}. His method was extended \cite{PMM:79,Mc:86} for regular arrays of circular cylinders. A method of functional equations \cite{Mi:97, Ryl:00} employing analytic functions was used to find an expression of the permittivity tensor for small volume fraction of inclusions. For rectangular lattice of inclusions an efficient method based on the use of elliptic functions was suggested in Ref.~\onlinecite{BK:01}. Numerical evaluation of complex-valued effective permittivity can be found in \cite{SKBB:96}, \cite{BB:99} as well as in \cite{M:04} where one can find the bounds on the transport coefficients of a two-dimensional composite.

The solution of the problem in question consists of two steps. First, we construct a quasiperiodic potential
using a combination of the Weierstrass $\zeta$-function and its derivatives. This ensures periodicity of the electric field in the whole plane and, as a result, avoids the problem of the summation of a conditionally convergent series. This approach is similar to that in Ref.~\onlinecite{GF:70} applied to biharmonic problems of the theory of elasticity. 
Second, we determine the average electric field and electric displacement within one parallelogram of periods and find an exact expression of the effective permittivity tensor that relates the two quantities. 
This paper is an extension of the author's previous article \cite{G:12} to the case when
physical properties of the component constituents are complex-valued. The use of complex
quantities allows us to greatly simplify the solution of the problem.

\section{Derivation of periodic potential}

\begin{figure}[h]
\includegraphics[width=0.4\textwidth, angle=0]{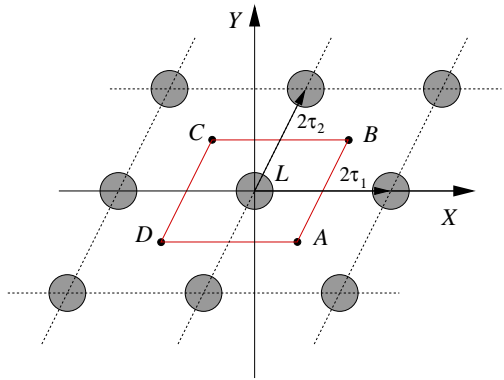}
\caption{\label{fig1} Circular inclusions of radius $a$ arranged in a
periodic lattice with periods $2\omo$ and $2\omt$.} 
\end{figure}

Suppose that a periodic lattice of identical circular inclusions of radius $a$
is introduced into a uniform complex electric field $E$ applied in the plane perpendicular to the cylinder axes. The nodes of the lattice in the complex plane are generated 
by a pair of vectors $2\omo$ and $2\omt$, $\im \frac{\omt}{\omo} > 0$ (see Figure~\ref{fig1}). 
In polar coordinates the complex-valued potential $u(\r,\te)$ has the following properties: 
\begin{align}
 \nabla \cdot [\bep(\r, \om) \nabla u(\r,\om)] =0, \quad u = \left\{
 \begin{array}{l}
  \ui \text{ in the inclusion}, \\[2mm]
  \ue \text{ in the medium},
 \end{array}
\right.
\end{align}
where $\bep(\r, \om) = \bep^\pr(\r, \om) + i \bep^\prr (\r,\om)$ is the tensor of complex permittivity at frequency $\om$. Hereafter prime and double prime denote the real and
imaginary part of the quantity, respectively.

Potentials $\ui$ and $\ue$ cannot be both analytic, and therefore we represent $u$ in the form 
\begin{align}
\label{Phi_in}
\ui (z) &= Ea \sum_{n=0}^{\infty} \left[ A_n \frin{z}{a}{2n+1}  
+ B_n \frin{\bar{z}}{a}{2n+1} \right], \quad |z| < a, \\
\ue (z) &= -E z + Ea \sum_{n=0}^{\infty} \frac{a^{2n+1}}{(2n)!} 
\left[ C_n \zeta^{(2n)}(z) + D_n  \zeta^{(2n)}(\bar{z})\right], \quad |z| > a,
\label{Phi_ex}
\end{align}
where $A_n, \; B_n,\; C_n,\;$ and $D_n$ are unknown complex dimensionless coefficients, 
${\bar z}$ stands for the complex conjugation, and $\zeta(z)$ is Weierstrass' $\zeta$-function 
\cite{BE:53}
\begin{equation}
\zeta (z) = \frac{1}{z} + {\sum_{m,n}}^{\prime} \left[ \frac{1}{z-\P} + \frac{1}{\P}
+ \frac{z}{\P^2} \right].
\label{zeta} 
\end{equation}
Here $\zeta^{(2n)}(z)$ denotes derivative of order $2n$, and $\P = 2m\omo + 2n\omt$.
Prime in the sum means that summation is extended over all pairs $m,\,n$ except $m=n=0$. 

On the boundary $r=a$ of inclusion we impose continuity conditions
\begin{align}
 & \left\llbracket u \right\rrbracket =0, \label{bc1} \\[2mm]
 & \left\llbracket \ga\, \frac{\de u}{\de r} \right\rrbracket = 0, \label{bc2}
\end{align}
where brackets $\llbracket \cdot \rrbracket$ denote the jump 
of the enclosed quantity across the interface.

\noindent
Below we will use some properties of $\zeta$-function \cite{AS:65,BE:53}.
It has the quasiperiodicity property
\begin{align}
 \label{q1}
 \zeta(z+2\tau_k) &= \zeta(z) + 2\eta_k, \quad \eta_k=\zeta(\tau_k),\quad k=1,2,
\end{align}
where constants $\eta_1$ and $\eta_2$ are related by the Legendre identity
\begin{equation}
 \eta_1 \,\omt - \eta_2 \,\omo = \frac{\pi i}{2}.
 \label{legendre}
\end{equation}
The derivative of $\zeta(z)$ is a periodic function and is expressed through 
Weierstrass elliptic function $\wp(z)$  by
\begin{equation}
 \zeta^\prime (z) = -\wp(z).
\end{equation}
This property ensures the electric field to be periodic in the medium, while
\rf{q1} guarantee that the potential changes by a constant value
in the direction of either $2\omo$ or $2\omt$.

To satisfy conditions \rf{bc1}-\rf{bc2} on the inclusion surface we expand $\zeta(z)$ 
and its even derivatives in a Laurent series 
\begin{align} 
 \zeta^{(2n)}(z) &= \frac{(2n)!}{z^{2n+1}}- \sum_{k=0}^\infty s_{n+k+1}\,
 \frac{(2n+2k+1)!}{(2k+1)!}\,z^{2k+1},
  \quad n \geq 0, \quad s_1=0,
\end{align}
where 
\begin{equation}
 s_k = {\sum_{n,m}}^\prime \frac{1}{\P^{\,2k}}, \quad k = 2,3, \ldots.
 \label{sk}
\end{equation}
Sums \rf{sk} contain only even powers of $\P$ since for every point
$\P = 2m\omo + 2n\omt$ on the lattice there exists symmetric point
$-\P$ and the sums with odd powers vanish. Also, if the periods $2\omo$ and
$2\omt$ of the lattice are fixed, sums $s_k$ remain bounded as $k \to \infty$. 

We introduce new variables
\begin{align}
 a_n &= A_n + B_n, \quad b_n = A_n -B_n, \\
 c_n &= C_n + D_n, \quad d_n = C_n - D_n.
\end{align}
In the new variables, compliance with conditions \rf{bc1}-\rf{bc2} leads to a system of equations
\begin{align}
\label{eq1}
 a_n &= \frac{2\ge}{\ge-\gi} c_n, \quad b_n = -\frac{2\ge}{\ge-\gi} d_n, \\[2mm]
 \label{eq2}
 c_n &-\alpha \sum_{m=0}^\infty \frac{(2n+2m+1)!}{(2n+1)!\,(2m)!}\,a^{2n+2m+2}\, s_{n+m+1}\,c_m =  \alpha \delta_{n,0}, \\[2mm]
 d_n &+\alpha \sum_{m=0}^\infty \frac{(2n+2m+1)!}{(2n+1)!\,(2m)!}\,a^{2n+2m+2}\, s_{n+m+1}\,d_m =  -\alpha \delta_{n,0}.
 \label{eq3}
\end{align}
where
\begin{equation}
  \alpha = \frac{\gi - \ge}{\gi + \ge}.
  \label{alpha}
\end{equation}
Using notation
\begin{align}
 \ell &= \min \{2|\omo|, 2|\omt|, 2|\omo - \omt| \}, \\
 h &=\frac{a}{\ell}, \quad h\leq\frac{1}{2}, \\
 S_k &= {\sum_{n,m}}^\prime \frin{\ell}{\P}{2k}, \; k=2,3,\ldots, \; S_1 = 0,
 \label{gk}
\end{align}
both equations \rf{eq2}-\rf{eq3} then can be written as
\begin{widetext}
\begin{align}
 x_n - \sum_{m=0}^\infty G_{n,m} \, x_m h^{2n+2m+2} = \alpha y \,\delta_{n,0},
 \label{vec_sys}
\end{align}
\end{widetext}
where
\begin{align}
 G_{n,m} = \pm \alpha\, \frac{(2n+2m+1)!}{(2n+1)!(2m)!}\, S_{n+m+1}, \quad
 G_{0,0} =0, \quad y = \mp 1.
 \label{Gnm}
\end{align}
The system \rf{vec_sys} has the same structure as that considered in \cite{G:12}. Therefore, its
unique solution exists for sufficiently small values of $h$ and can be obtained in the form
of a convergent power series
\begin{equation}
 x_n = \alpha y \,\delta_{n,0} + \sum_{m=0}^\infty p_{n,m} h^{2n+2m+2},
 \label{ser}
\end{equation}
where
\begin{align}
 p_{n,0} &= \alpha G_{n,0}\, y, \nonu \\
 p_{n,k} &= \sum_{m=0}^{\left[ \frac{k-1}{2} \right]} G_{n,m}\, p_{m,k-2m-1},\quad k = 1,2,\ldots.
 \label{pnk}
\end{align}
Calculation of $c_0$ and $d_0$ from \rf{ser}--\rf{pnk} gives  
\begin{equation}
 c_0 = \alpha \lambda, \quad d_0 = -\alpha \mu,
 \label{c0}
\end{equation}
where
\begin{align}
 \label{lambda}
 \lambda &=  1 + 3\alpha^2 S_2^2 h^8 + 5\alpha^2 S_3^2 h^{12} +30\alpha^3 S_2^2 S_3 h^{14}
 +\left(9\alpha^4 S_2^4 + 7\alpha^2 S_4^2 \right)h^{16} \nonu \\
 &+210\alpha^3 S_2 S_3 S_4 h^{18} + (330\alpha^4 S_2^2 S_3^2 + 9\alpha^2 S_5^2 )h^{20} +O(h^{22}), \\
 \mu &= 1 + 3\alpha^2 S_2^2 h^8 + 5\alpha^2 S_3^2 h^{12} -30\alpha^3 S_2^2 S_3 h^{14}
 +\left(9\alpha^4 S_2^4 + 7\alpha^2 S_4^2 \right)h^{16} \nonu \\
 &-210\alpha^3 S_2 S_3 S_4 h^{18} + (330\alpha^4 S_2^2 S_3^2 + 9\alpha^2 S_5^2 )h^{20} +O(h^{22}). 
 \label{mu}
\end{align}

\section{Calculation of the effective permittivity tensor}

Effective permittivity tensor ${\bm \gamma}^{\ast}$
relates the average electric displacement $\langle {\bm D} \rangle$ and the average electric field $\langle \BE \rangle$
\begin{equation}
\langle{\Bj}\rangle = {\bm \gamma}^{\ast} \langle{\bm E}\rangle.
\label{j}
\end{equation}
Observe that
\begin{align}
 \la{\Bj}\ra &= \frac{1}{S} \iint_S \Bj\,dS 
 =  \frac{\gamma_{in}}{S} \iint_{S_{in}} \BE_{in}\,dS + \frac{\se}{S} \iint_{S_{ex}} \BE_{ex}\,dS.
\end{align}
Calculation of the above integrals gives \cite{G:12}
\begin{align}
 \label{Ein}
 \frac{1}{S} \iint_{S_{in}} \BE_{in}\,dS & = 
 \frac{2\pi a^2 \se}{(\gamma_{in} + \se)S}\,\M \BE, \\[2mm]
 \frac{1}{S} \iint_{S_{ex}} \BE_{ex}\,dS & =\BE - \frac{2 \alpha a^2}{S}\, {\bm \Psi}
 \M \BE
 \label{Eex}
\end{align}
where
\begin{align}
 \M = \left[
 \begin{array}{cc}
  \lambda & 0 \\[-1ex]
  0 & \mu
 \end{array}
\right], \quad
\BE = E \J,
\label{M}
\end{align}
and matrix ${\bm \Psi}$ depends only on geometry of the lattice
\begin{align}
 {\bm \Psi} &= \left[
 \begin{array}{cc}
 \re \eta_1 \im 2\omt & -\im \eta_1 \im 2\omt \\
 -\im \eta_1 \im 2\omt & \pi - \re \eta_1 \im 2\omt
 \end{array}
 \right].
 \label{Psi}
\end{align}
Computing the average electric field $\langle \BE \rangle$ we obtain
\begin{equation}
 \la \BE \ra = \left( \I - \frac{2\alpha a^2}{S}\, {\bm \Psi} \M \right)\E.
 \label{Eav}
\end{equation}
From here and \rf{Ein}--\rf{Eex} we derive the effective permittivity tensor
\begin{equation}
 \Bs^\ast = 
 \ge \left\{\I + \pi \delta \M
\left( \I - \delta {\bm \Psi} \M
 \right)^{-1}
\right\}, 
\label{eps}
\end{equation}
where
\begin{align}
 \delta  = \frac{2\alpha a^2}{S}.
 \label{delta}
\end{align}
Explicit computation of the $2\times 2$ inverse matrix in \rf{eps} shows that $\Bs^\ast$ is symmetric.
If $\| \delta {\bm \Psi} \M\| < 1$ then $\Bs^\ast$ can be
expanded in a convergent series
\begin{align}
 \Bs^\ast 
 &=\ge \I+  \ge \pi \delta  \M \sum_{n=0}^\infty \left( \delta {\bm \Psi} \M\right)^{n}.
\label{sig2}
\end{align}
In what follows we use expression \rf{eps} for calculation of the effective permittivity tensor for specific lattices.

\section{Examples}

\subsection{Square lattice}
\begin{figure}[h]
\includegraphics[width=0.35\textwidth, angle=0]{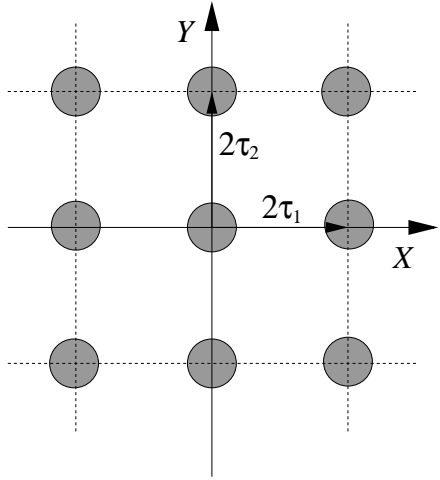}
\caption{Square lattice of inclusions of radii $a$ with periods $2\omo=\ell$ and $2\omt=i\ell$.} 
\label{fig2}
\end{figure}

\begin{figure}[h]
\includegraphics[width=0.45\textwidth, angle=0]{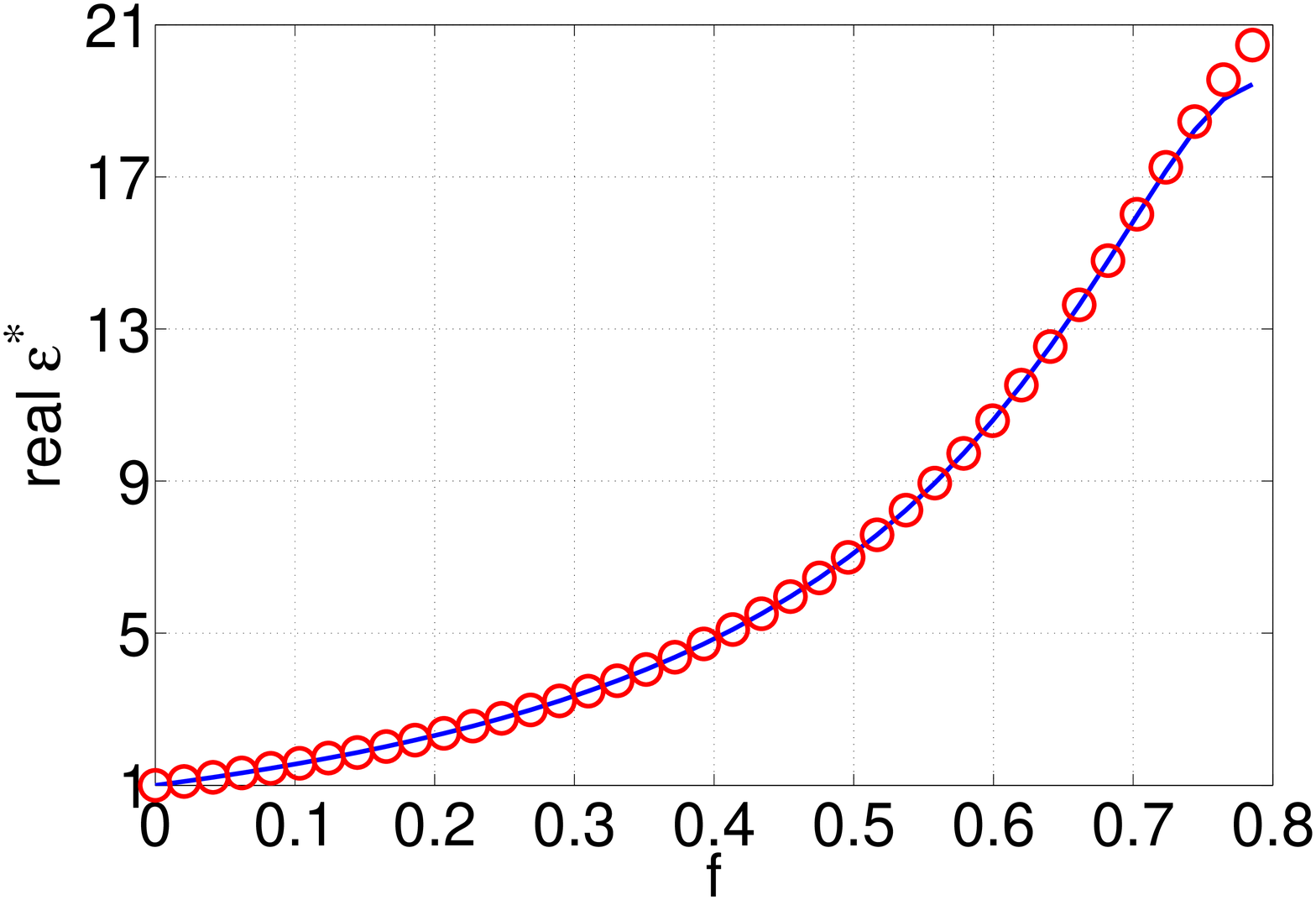} 
\includegraphics[width=0.45\textwidth, angle=0]{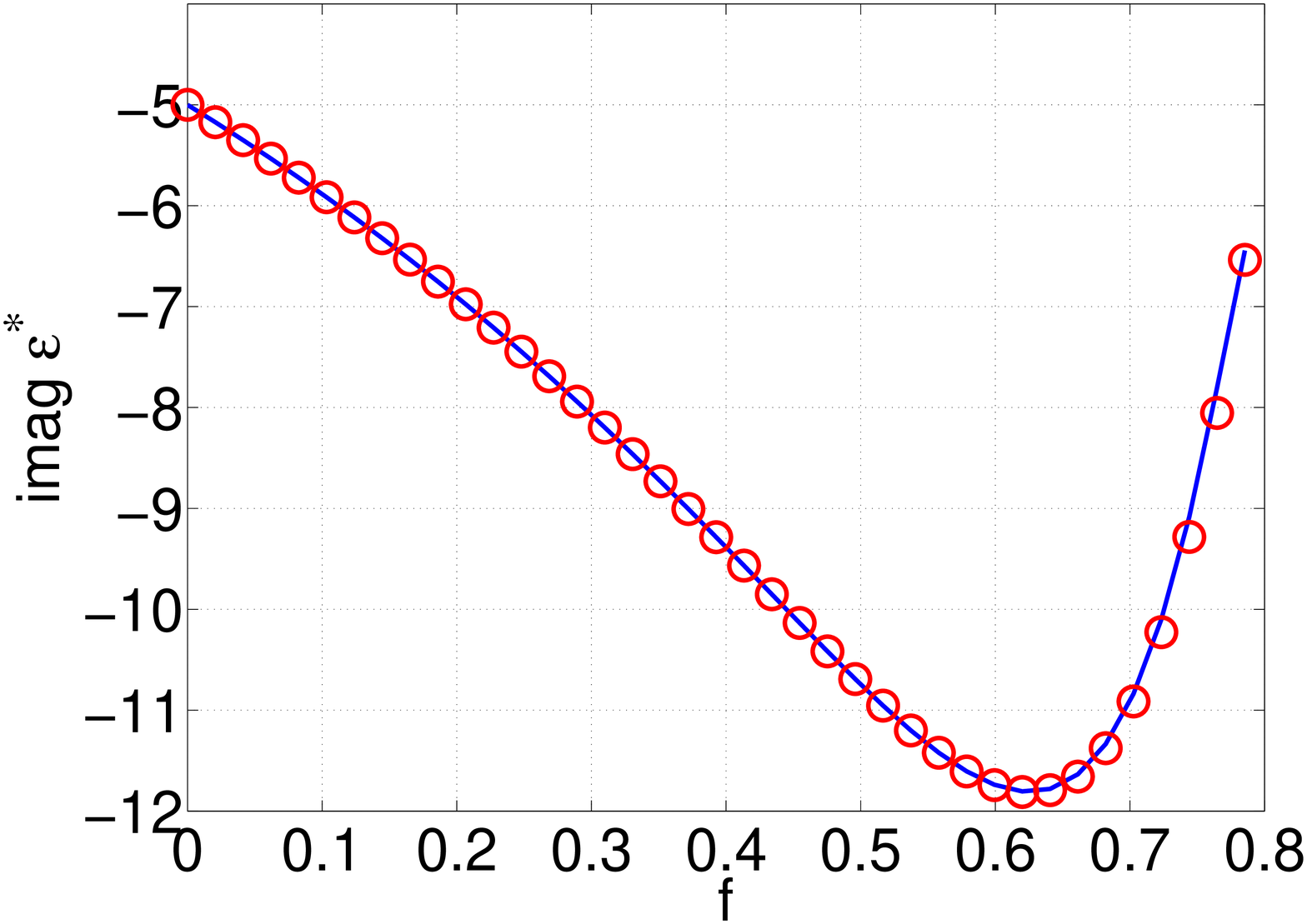}\\
(a) \hspace*{70mm}(b)
\caption{Dependence of the real (a) and imaginary (b) part of the complex effective dielectric constant $\ep^\ast$ on the volume fraction $f$ evaluated numerically (solid blue line) and by formula \rf{ep_iso} (red circles) for a square array of cylinders. We used $\ep_{in}=30-0.3i$ and $\ep_{ex}=1-5i$.} 
\label{square1}
\end{figure}

\begin{figure}[h]
\includegraphics[width=0.45\textwidth, angle=0]{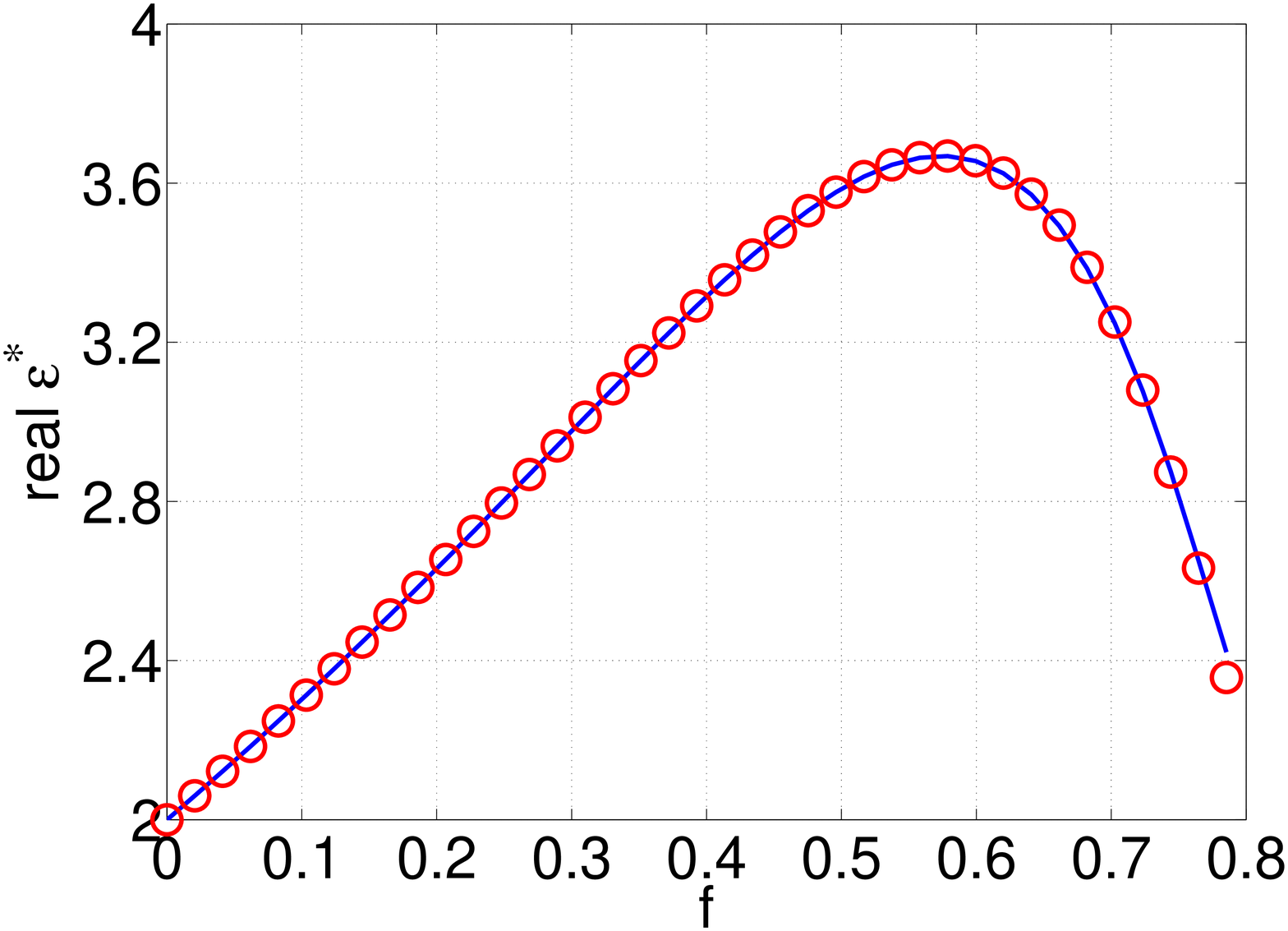} 
\includegraphics[width=0.45\textwidth, angle=0]{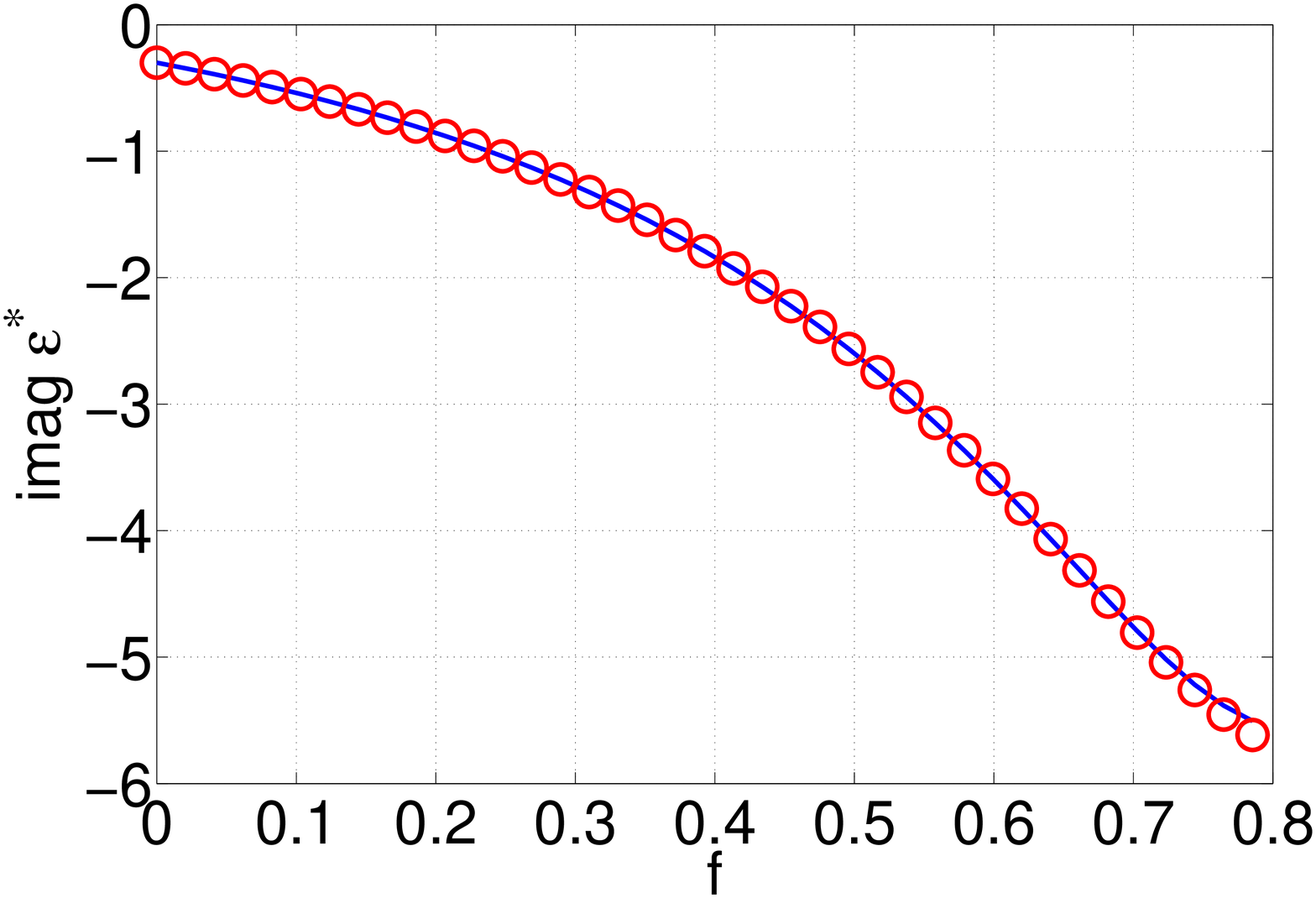}\\
(a) \hspace*{70mm}(b)
\caption{Dependence of the real (a) and imaginary (b) part of the complex effective dielectric constant $\ep^\ast$ on the volume fraction $f$ evaluated numerically (solid blue line)
and by formula \rf{ep_iso} (red circles) for a square array of cylinders when $\ep_{in}=1-8i$ and $\ep_{ex}=2-0.3i$.} 
\label{square2}
\end{figure}

For the square lattice we put $2\omo = \ell,\; 2\omt=i\ell$ (see Figure \ref{fig2}). 
Then one can find \cite{AS:65}  that $\dst \eta_1 = \frac{\pi}{2\ell}$.
From \rf{Psi} we obtain $\dst {\bm \Psi} = \frac{\pi}{2} \I$.
All lattice sums \rf{gk} are real with the only non-zero being $S_{2k}$, $k=1,2,\ldots$.
Calculation of matrix $\M$ in \rf{M} gives $\lambda = \mu$. Hence tensor
$\Bs^\ast = \ep^\ast \I$ is isotropic with
\begin{align}
 \ep^\ast &= \ep_{ex}\,\frac{1+\alpha \lambda f}{1-\alpha \lambda f},
 \label{ep_iso}
\end{align}
where
\begin{align}
 \lambda &=  1 + 3\alpha^2 S_2^2 h^8 + \alpha^2 \left(9\alpha^2 S_2^4 + 7 S_4^2 \right)h^{16} + \left( 27\alpha^6 S_2^6 + 21\alpha^4 S_2^2 S_4^2 + 2205\alpha^4 S_2^2 S_4^2 \right. \nonu \\
 &+ \left.21\alpha^4 S_2^2 S_4^2 +11\alpha^2 S_6^2 \right)h^{24} + O(h^{32}).
 \label{lam_square}
\end{align}
Here $\dst f=\pi h^2$ is the volume fraction of the inclusions, $\; \dst h=\frac{a}{\ell}$,
$\dst S_2 = {\sum_{n,m}}^\prime \frac{1}{(m+in)^4} = 3.15121$, $\dst S_4 = {\sum_{n,m}}^\prime \frac{1}{(m+in)^8}=4.25577$, $\dst S_6 = {\sum_{n,m}}^\prime \frac{1}{(m+in)^{12}} = 3.93885$ correct to five decimal places. Dependence of the real and imaginary parts of $\ep^\ast$ on the volume fraction of the cylinders is shown in Figures \ref{square1} and \ref{square2}
for $\ep_{in}=30-0.3i$, $\ep_{ex}=1-5i$ and $\ep_{in}=1-8i$, $\ep_{ex}=2-0.3i$, respectively.
In both examples the dielectric constants are chosen in such a way to make either real or
imaginary part of $\ep^\ast$ nonmonotonic. Formula \rf{ep_iso} gives an excellent agreement between numerical evaluation of $\ep^\ast$ form \rf{eq2}, \rf{c0} and calculation by \rf{lam_square}. 

\subsection{Regular triangular lattice}

\begin{figure}[h]
\includegraphics[width=0.55\textwidth, angle=0]{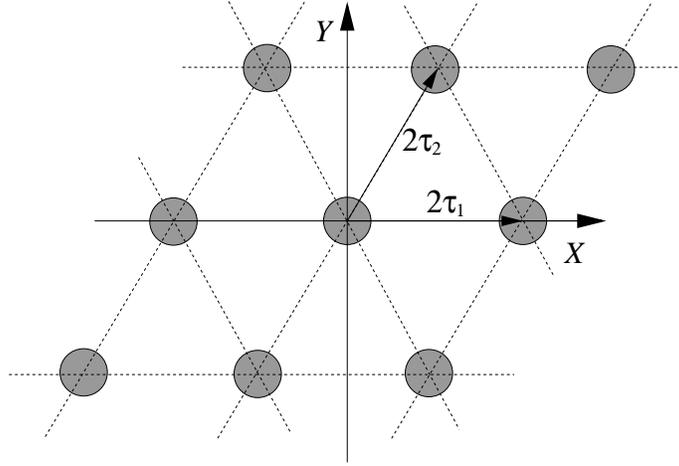}
\caption{Regular triangular lattice of inclusions of radii $a$ with periods $2\omo=\ell$ and $2\omt=\ell e^{i\pi/3}$.} 
\label{fig3}
\end{figure}

\begin{figure}[h]
\includegraphics[width=0.45\textwidth, angle=0]{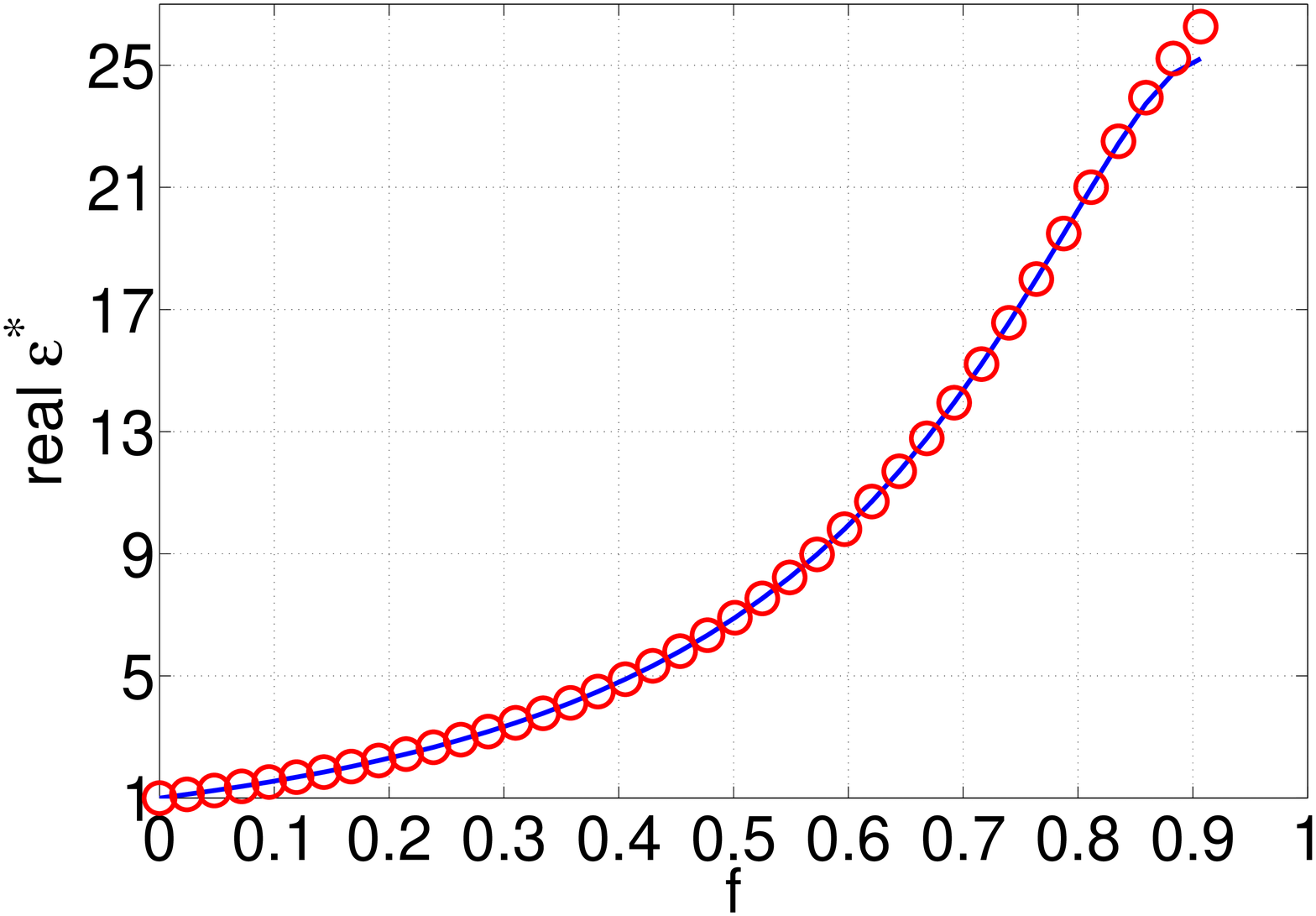} 
\includegraphics[width=0.45\textwidth, angle=0]{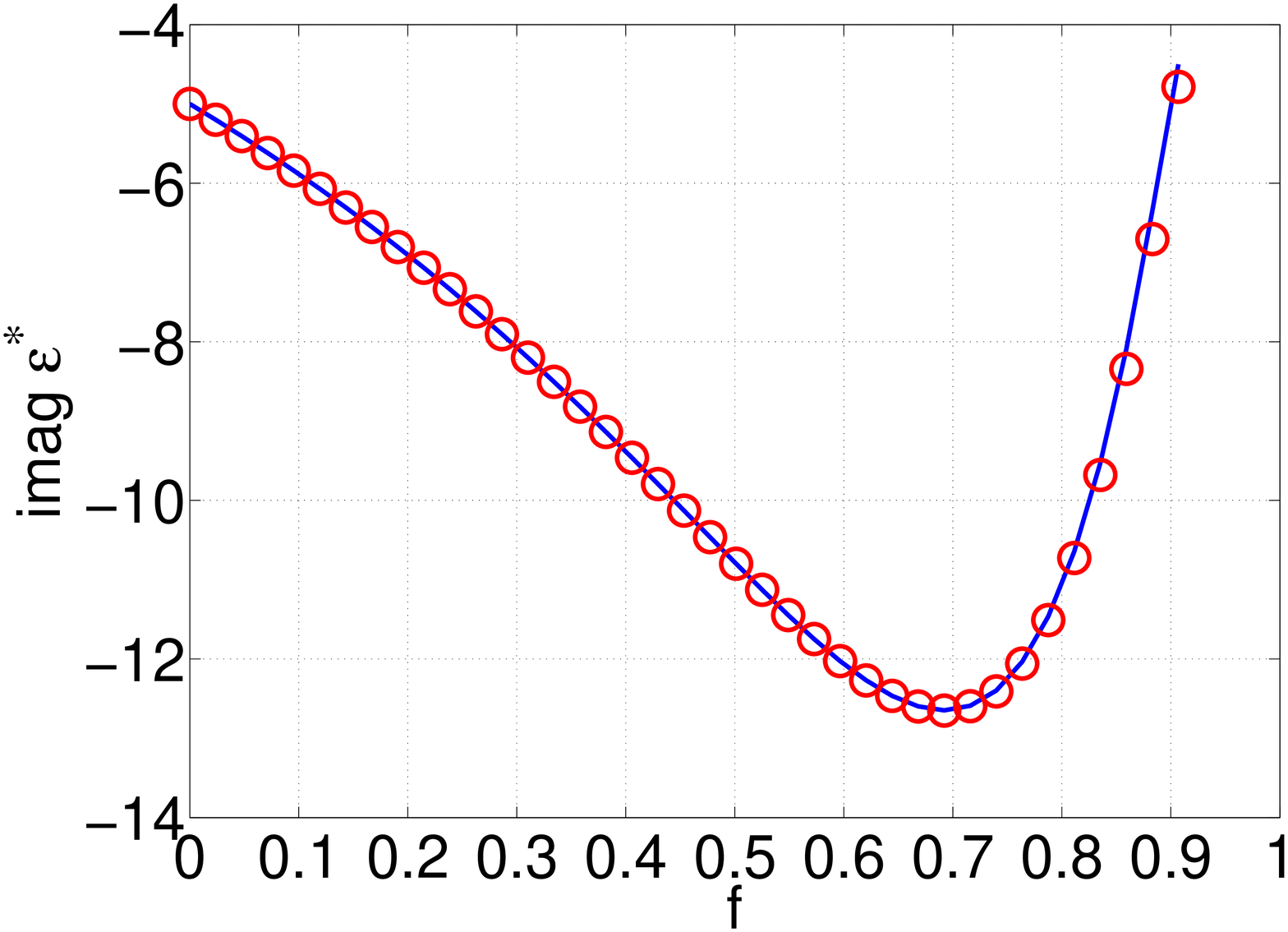}\\
(a) \hspace*{70mm}(b)
\caption{Dependence of the real (a) and imaginary (b) part of the complex effective dielectric constant $\ep^\ast$ of a regular triangular array of cylinders on their volume fraction $f$  evaluated numerically (solid blue line)
and by formula \rf{ep_iso} (red circles) for $\ep_{in}=30-0.3i$ and $\ep_{ex}=1-5i$.} 
\label{triang1}
\end{figure}

\begin{figure}[h]
\includegraphics[width=0.45\textwidth, angle=0]{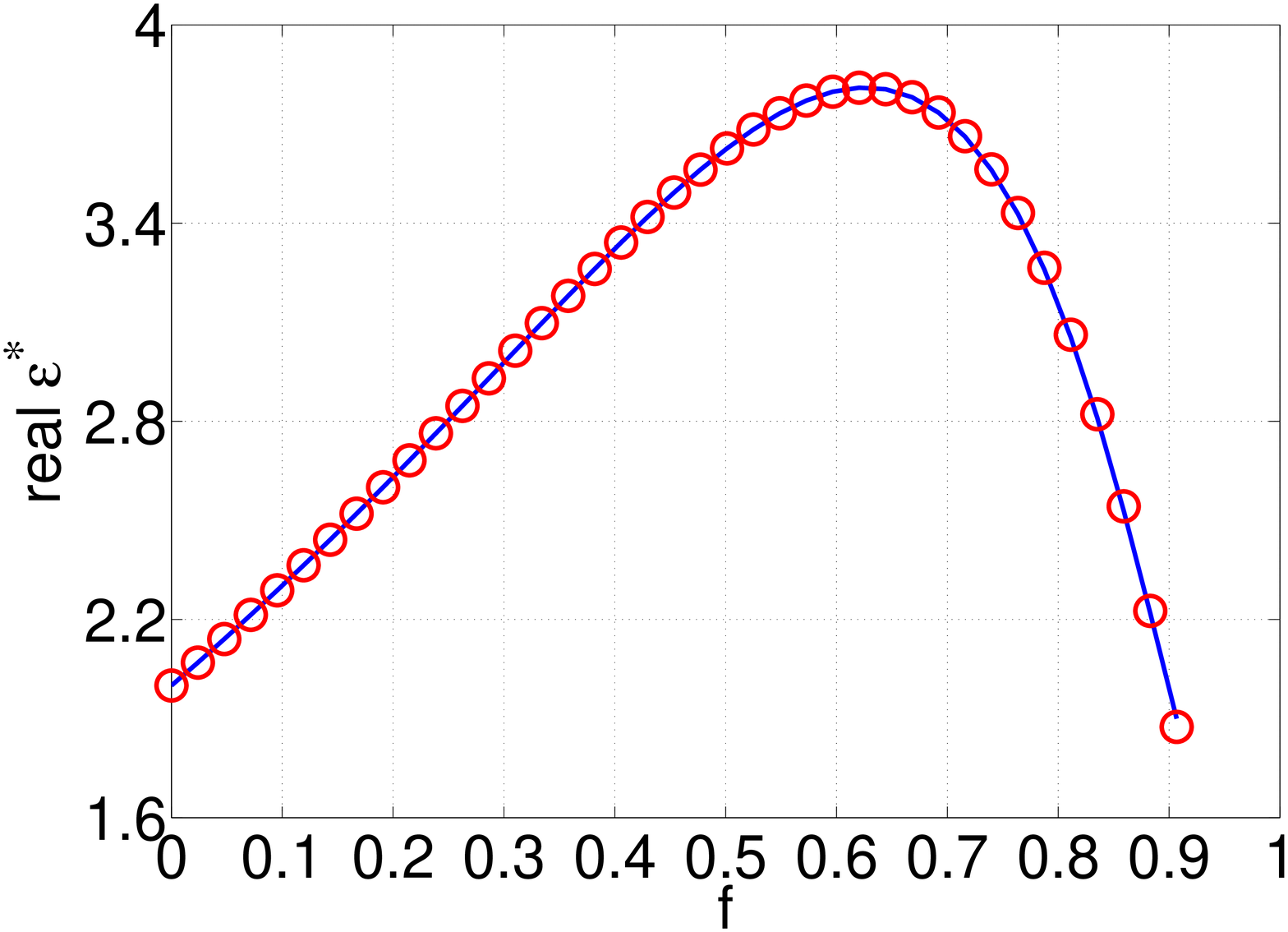} 
\includegraphics[width=0.45\textwidth, angle=0]{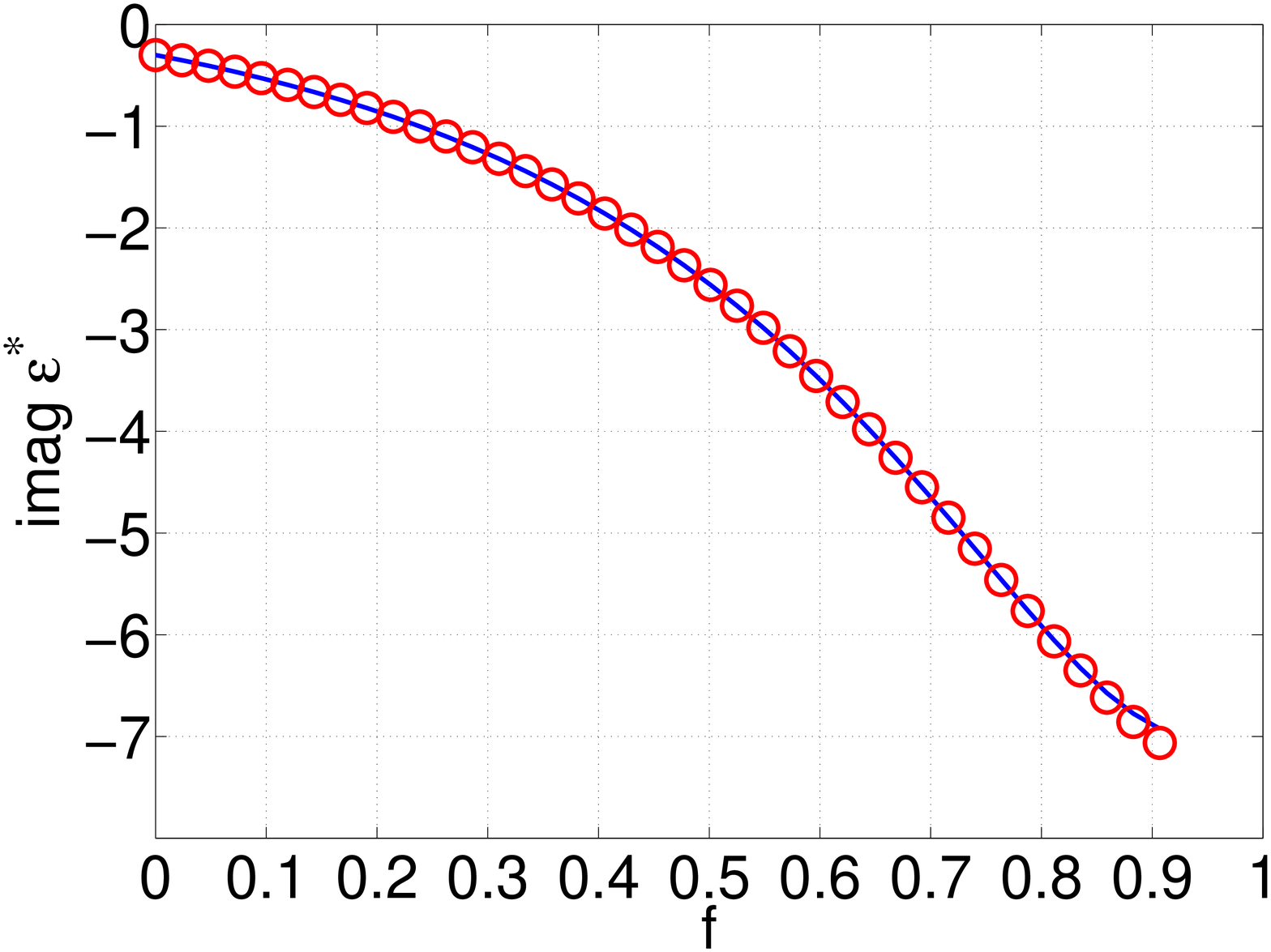}\\
(a) \hspace*{70mm}(b)
\caption{Dependence of the real (a) and imaginary (b) part of the complex effective dielectric constant $\ep^\ast$ of a regular triangular array of cylinders on their volume fraction $f$ evaluated numerically (solid blue line)
and by formula \rf{ep_iso} (red circles) for $\ep_{in}=1-8i$ and $\ep_{ex}=2-0.3i$.} 
\label{triang2}
\end{figure}

For triangular lattice (see Figure \ref{fig3})  we put $\dst 2\omo = \ell,\; 2\omt=\ell e^{\pi i/3}$. 
Then we find \cite{AS:65} the constant $\dst \eta_1 = \frac{\pi}{\ell \sqrt{3}}$.
As a result, matrices $\dst {\bm \Psi}=\frac{\pi}{2} \I$ and $\M = \lambda \I$
are isotropic as well as $\Bs^\ast = \ep^\ast \I$. We can use \rf{ep_iso} with
\begin{equation}
 \lambda = 1 +5\alpha^2 S_3^2 h^{12} +\alpha^2(25\alpha^2 S_3^4 +11S_6^2)h^{24}
 + O(h^{36}),
 \label{lambda_tri}
\end{equation}
where $\dst f = \frac{2\pi}{\sqrt{3}}h^2,\;h=\frac{a}{\ell}$, while the only non-zero real
lattice sums are $S_{3k}$, $k=1,2,\ldots$.
Here $\dst S_3 = {\sum_{n,m}}^\prime \frac{1}{\left(m+ne^\frac{\pi i}{3}\right)^6}=5.86303$,
$\dst S_6 = {\sum_{n,m}}^\prime \frac{1}{\left(m+ne^\frac{\pi i}{3}\right)^{12}}=6.00964$
correct to five decimal places. Dependence of the real and imaginary parts of $\ep^\ast$ on the volume fraction of the cylinders is shown in Figures \ref{triang1} and \ref{triang2}
for $\ep_{in}=30-0.3i$, $\ep_{ex}=1-5i$ and $\ep_{in}=1-8i$, $\ep_{ex}=2-0.3i$, respectively.
In both examples the dielectric constants are chosen in such a way to make either real or
imaginary part of $\ep^\ast$ nonmonotonic. Formula \rf{ep_iso} gives an excellent agreement between numerical evaluation of $\ep^\ast$ form \rf{eq2}, \rf{c0} and calculation by \rf{lam_square}. 

\subsection{Anisotropic lattice}

As an example of anisotropic lattice we consider the latice with the periods 
$2\omo = \ell$, $2\omt = \ell e^{3\pi i/8}$ (see Figure \ref{anisotr}). Then the constant $\eta_1$ can be found numerically  $\eta_1 = \zeta(\omo)
\approx (1.73289 -0.07888i)/\ell$, and matrix ${\bm \Psi}$ in \rf{Psi} becomes
\begin{equation}
 {\bm \Psi} = \left[
 \begin{array}{rr}
  3.20196 & 0.14575 \\
  0.14575 & -0.06037  
 \end{array}
\right].
\end{equation}
Lattice sums for approximation of $\lambda$ and $\mu$ in \rf{lambda}-\rf{mu} have the values
$\dst S(2) = {\sum_{n,m}}^\prime \frac{1}{\left(m+ne^\frac{3\pi i}{8}\right)^4}=1.01011 + 1.01011i$, $\dst S(3) = {\sum_{n,m}}^\prime \frac{1}{\left(m+ne^\frac{3\pi i}{8}\right)^6}= 4.28856 - 1.77638i$, $\dst S(4) = {\sum_{n,m}}^\prime \frac{1}{\left(m+ne^\frac{3\pi i}{8}\right)^8}= 0.87457i$, $\dst S(5) = {\sum_{n,m}}^\prime \frac{1}{\left(m+ne^\frac{3\pi i}{8}\right)^{10}}=2.78468 + 1.15345i$. Then we determine components of the effective dielectric tensor from \rf{eps} using the solution of \rf{eq2}-\rf{eq3} for numerical evaluation of 
matrix $\M$ and its approximation by \rf{lambda}-\rf{mu}. Figure \rf{anisotr} shows
dependence of the components of the real and imaginary parts of $\bep^\ast$ on the volume 
fraction $f$ of the cylinders. Lines denote component of the effective tensor obtained from numerical solution of \rf{eq2}--\rf{eq3}, while symbols correspond to their approximation by \rf{lambda}--\rf{mu}. As in the previous cases, agreement between numerical evaluation and analytic approximation of the effective dielectric tensor is excellent even for
high values of $f$.
\begin{figure}[h]
\includegraphics[width=0.5\textwidth, angle=0]{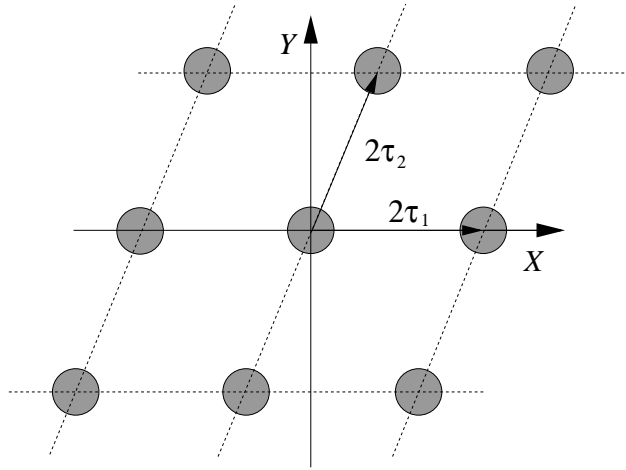}
\caption{Anisotropic lattice of inclusions of radii $a$ with periods $2\omo=\ell$ and $2\omt=\ell e^{3\pi i/8}$.} 
\label{fig4}
\end{figure}

\begin{figure}[h]
\includegraphics[width=0.45\textwidth, angle=0]{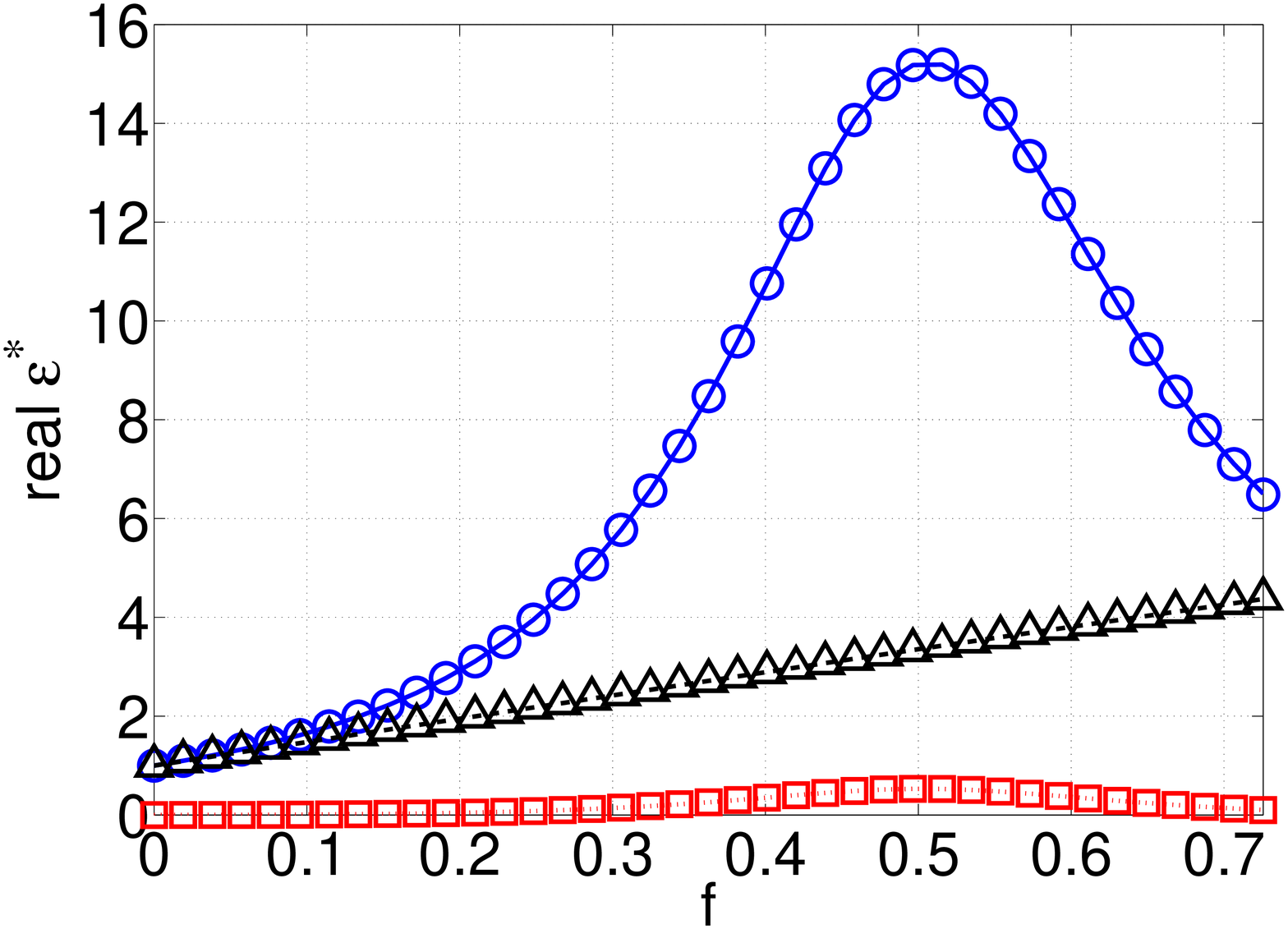} 
\includegraphics[width=0.45\textwidth, angle=0]{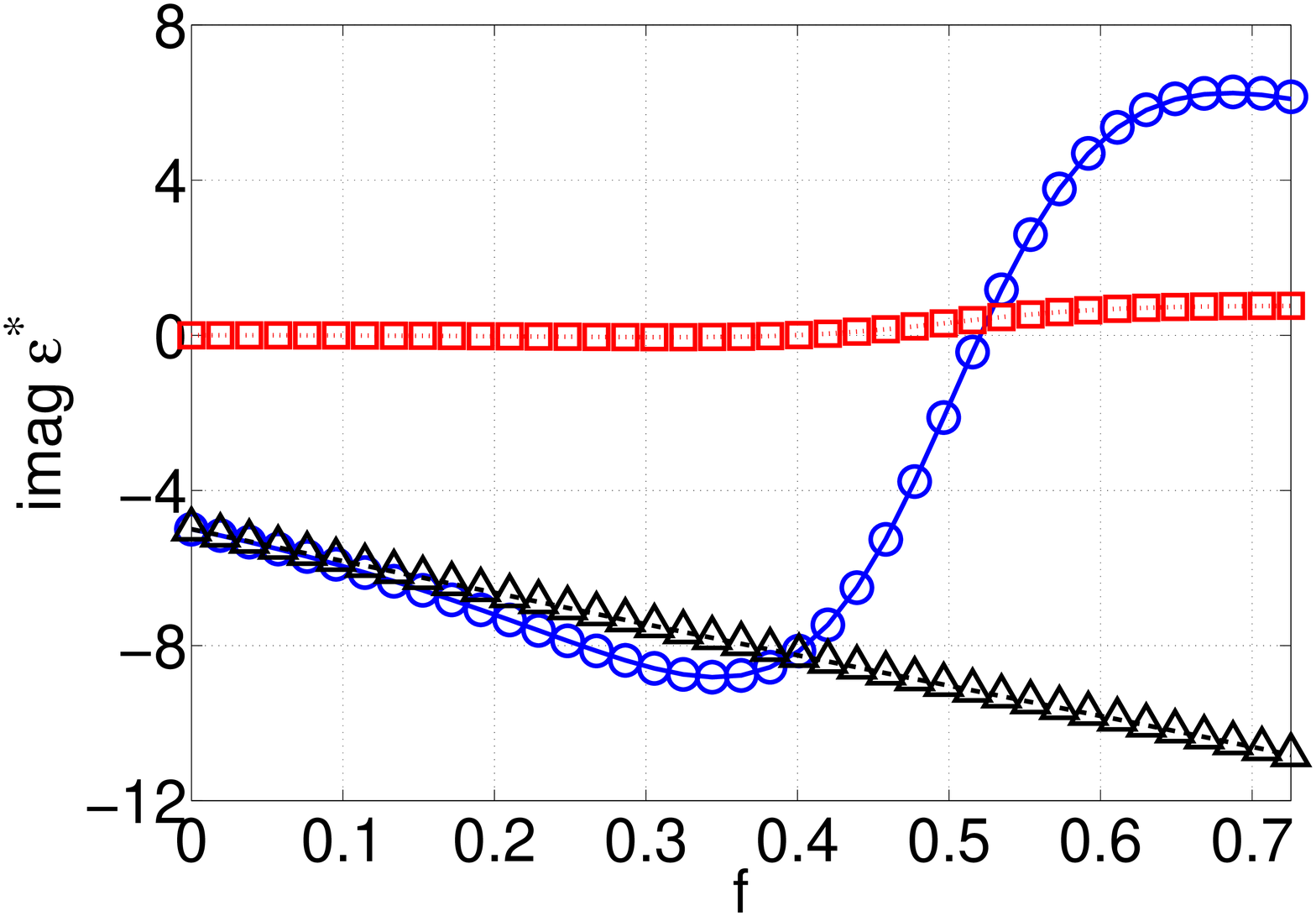}\\
(a) \hspace*{70mm}(b)
\caption{Dependence of the real (a) and imaginary (b) part of the complex effective dielectric tensor $\bep^\ast$ of a lattice with periods $2\omo = \ell$, $2\omt = \ell e^{3\pi i/8}$ on the volume fraction $f$ of the cylinders. The solid blue line is $\ep^\ast_{11}$, the dashed black line
corresponds to $\ep^\ast_{22}$, the dotted red line is $\ep^\ast_{12}$ evaluated numerically. Blue circles, black triangles, and red squares correspond to their approximation by \rf{lambda}-\rf{mu}, respectively. The dielectric constants of the materials are $\ep_{in}=30-0.3i$ and $\ep_{ex}=1-5i$.} 
\label{anisotr}
\end{figure}

\section{Conclusion}

We determine the effective complex permittivity of a two-dimensional composite,
consisting of an arbitrary doubly periodic array of identical circular cylinders in a homogeneous matrix, and whose dielectric properties are complex-valued. Efficient formulas are provided to determine the effective complex permittivity tensor which are in excellent agreement with numerical calculations in the whole range of cylinder volume fraction variation. We also show that in contrast to the real-valued case, the real and imaginary parts of the effective complex-valued tensor can exhibit non-monotonic behavior as functions of volume fraction of cylinders, and can be either greater or less than that of the constituents. This raises the question about optimal design of such composites.

\end{document}